\documentstyle[12pt]{article}
\begin{document}
\vspace{5cm}
\begin{center}

{\Large \bf { Nonperturbative effects \\
in the nonrelativistic hadron scattering}}

\vspace{1cm}

{\it {V.I.Shevchenko\footnote{e-mail:shevchenko@vitep3.itep.ru} }}

\bigskip

{\bf {Institute of Theoretical and Experimental Physics,\\}}

{ {B.Cheremushkinskaya, 25, Moscow, Russia}}

\vspace{3cm}

{\bf\it {Abstract}}
\end{center}

\bigskip

Starting from the gauge-invariant quark-antiquark Green function
the expression for the nonrelativistic meson scattering amplitude
on the external gluon field is obtained. The nonperturbative form of this amplitude is
discussed. It is shown, that nonlocality of the nonperturbative vacuum field
distribution plays an essential role. As a concrete example, charmonium-nucleon
scattering length is found and the possibility of the exotic charmonium-nucleus
bound states is analyzed.

\newpage
{\hspace{5mm} The structure} and properties of the vacuum are among the most
interesting questions of modern quantum theory. Of particular interest is
the vacuum in nonabelian gauge theories. There are good grounds to believe,
that this structure is very rich and complicated. Attempts to understand the
hadron properties on the basis of the vacuum ones make an integral part of
all papers in hadron physics.

This paper is mainly devoted to the meson-meson scattering in QCD vacuum.
In perturbative QCD quarks interact with each other by
exchange of gluons and the role playing by one or other Feynman diagram
depends on the kinematic region.
It was also proposed a nonperturbative model for
the scattering in the Regge limit $ {s\gg m^2} $,
$ {t\sim m^2} $, where $m^2$ is a generic mass of the hadron
(see papers \cite{land}, \cite{nacht}, \cite{dosch2}) where
quark-(anti)quark interaction is the result of QCD vacuum nontriviality
and corresponding amplitude is proportional to the nonperturbative vacuum
average of the gluon field stength 4-point function.

Here and below we are going to describe the (nonperturbative) effects, which
arise due to the interaction between nonrelativistic colour objects and nonperturbative gluon fields
fluctuations in QCD vacuum in terms of the so called method of vacuum correlators
(MVC) {\cite{simonov88}}, {\cite{dosch1}}.
The corresponding physical picture in the case of nonrelativistic
scattering is close to the one, developed in \cite{nacht}, \cite{dosch2},
but there are some differences.
First, for the ultrarelativistic Regge scattering the
typical distances in the impact parameter plane are quite large and this is a
reason to consider 'soft' Pomeron as a nonperturbative object. At the same time
its structure may be more complicated, than it is proposed in (\cite{land} -
\cite{dosch2}), in particular, it may include bound states, propagated in
$t$-channel.
Here we are going to
discuss the role nonperturbative vacuum fields play in the heavy-quarkonium
nonrelativistic scattering and to analyze the possibility of existence the
exotic charmonium bound state in nuclei,  which was suggested in {\cite{br}}.
In this case we are interested in point-like type of interaction because of
small quarkonium radius and that's
why we can apply the framework, which was discussed in \cite{land} - \cite{dosch2}.
Second, it should
be noticed, that we are looking for the scattering length for relatively
slow heavy particle, which is propagated through the heavy nucleus.
So we are going to explore usual nonrelativistic
expansion over $v/c$, not the eikonal approximation, as it was done in
(\cite{land} - \cite{dosch2}), and to restrict ourselves by the so called
bilocal correlator, as one often does in the stochastic vacuum model
\cite{simonov88}. The resulting expression for the amplitude corrects the
formula, suggested in \cite{peskin} and provides the clear way for
including the nonperturbative effects.


{\hspace{5mm}}A reasonable basis for the analysis
may be, in our opinion, the hadron-hadron scattering description in terms
of the connected Green function (G.f.) cluster expansion. As the
first and the simplest step we are going to discuss meson G.f. only
in this paper.

As it was noticed, our technical basis is vacuum correlators formalism.
In MVC vacuum is characterized by a set of cumulants like:  $$
{<F_{\mu \nu}(x_1) \Phi(x_1,x_2)F_{\rho \phi}(x_2)...  \Phi(x_{n-1},x_{n})
F_{\gamma \delta}(x_n) \Phi(x_n,x_1)\> >}_A \leqno (*) $$ where $ F_{\mu
\nu}(x) $ is the gluon field strength, $ \Phi(x_1,x_2) = P\exp ig
\int\limits_{x_2}^{x_1} A_{\mu}dz^{\mu} $ is the parallel transporter and $
n=1,2,...$ .  Also $ <...>_A $ means the average over all field configurations
with the standard (euclidean) mesure.


Consider the gauge-invariant Green function for a meson in an external colour
field.
It is easy to show that in bilocal approximation \cite{quarkonia}:

\begin{eqnarray}
&G_2^A(z_1,z_2) = e^{-m(T_2 - T _1)} \cdot \tilde G_1^0 ( {\vec R}_1(T_1);\>
 {\vec R}_2(T_2) )\;\cdot \tilde G_1^{V_1}( \vec r_1({T_1});\>\vec
 r_2({T_2})) + \nonumber \\
  {}&+ (4m\mu)^2 \frac{(ig)^2}{2N_c}
\int\limits_{C}^{}dz_{\mu}\int\limits_{C}^{}dz^{\prime}_{\nu}
<b^a_{\mu}(z)\>b^a_{\nu}(z^{\prime})>_A \; \int d^3 \vec R ({\tau}) \int d^3
\vec R ({\tau}^{\prime}) \nonumber \\
&\int d^3\vec r(\tau) \int d^3\vec r({\tau}^{\prime})
 e^{-m(\tau - T _1)} \cdot \tilde G_1^0 ( {\vec R}_1(T_1); {\vec
 R}(\tau) )\;\cdot \tilde G_1^{V_1}(\vec r_1({T_1});\>\vec
r(\tau))\cdot\nonumber \\&\cdot e^{-m({\tau}^{\prime} - \tau)} \cdot \tilde
 G_1^0 ( {\vec R}(\tau); {\vec R}({\tau}^{\prime}) )\;\cdot \tilde
G_1^{V_8}(\vec r(\tau);\>{\vec r\>}'({\tau}'))\cdot \nonumber\\ &\cdot
e^{-m(T_2 - {\tau}^{\prime})} \cdot \tilde G_1^0 ( {\vec
R}({\tau}^{\prime}); {\vec R}_2(T_2) )\;\cdot \tilde G_1^{V_1}({\vec
r\>}'({\tau}');\>\vec r_2(T_2)) \end{eqnarray}
Here $ Tr(b_{\mu}b_{\nu}) =
\frac12\; b^a_{\mu}b^a_{\nu} $
and nonrelativistic G.f. :  $$ \tilde G_1^{V}(\vec r_1({\tau}_1);\>\vec
r_2({\tau}_2)) =   \frac{1}{2\mu} \int\limits_{\vec r ({\tau}_1)=\vec
  r_1}^{\vec r ({\tau}_2)= \vec r_2}\>D\vec r\;
e^{-\frac{\mu}{2}\int\limits_{{\tau}_1}^{ {\tau}_2} \left ( \frac{d\vec
 r}{d{\tau}} \right ) ^2 d{\tau} +
 \int\limits_{{\tau}_1}^{{\tau}_2} Vd{\tau}}
$$
The $b^a_{\mu}$ here are the external gluon fields, may be nonperturbative.
The above Green function containes, up to the approximations have been made,
all information about meson dynamics in the external field. In
\cite{quarkonia} this expression had been averaged over all vacuum fields and
after it nonperturbative energy level shifts were extracted. We will understand
these fields as belonging to the nonperturbative source, nucleon in our
case, and the average over the gluon field states of this source will be
made.  The potentials $V_1$ and $V_8$ are the ones in the singlet and octet
meson state respectively.  We are going to analyze heavy quarkonium
scattering, so for $V_1$ and $V_8$ we limit ourselves by pure perturbative
expressions:

$$
{ V_1 = \frac{C_F
{\alpha}_S}{r}\;}
 ; \;{V_8 = -\frac{{\alpha}_S}{2N_c}\frac1r} ;
 \;C_F=\frac43\>;\;N_c=3.
$$
Notice, $ {\alpha}_s $ is renormalized here on the
characteristic quarkonia sizes - see discussion in {\cite{rig}}.

It is convenient to express $ G_2^A $  in terms of fields operators. The simplest
way to do it is by nonabelian Stokes theorem
{\cite{dosch1}}:
\begin{eqnarray*}
\int\limits_Cdz_{\mu}\>\int\limits_Cdz_{\nu}'\; b_{\mu}(z)\>b_{\nu}(z') =
\int d{\sigma}_{\mu\rho}(w)\>\int d{\sigma}_{\nu\lambda}(w')\cdot \\
\cdot\left ( \Phi(x_0,w)\>G_{\mu\rho}(w)\>\Phi(w,x_0)\cdot\Phi(x_0,w')\>
G_{\nu\lambda}(w')\>\Phi(w',x_0)\right )
\end{eqnarray*}
Here $ d{\sigma}_{\mu\nu} = a_{\mu\nu}d\tau d\beta\; ;
\;G_{\mu\rho}(w) $ - gluon field tensor.
The surface is parametrized by straight-line ans{$\ddot{\mbox a}$}tze: $\;
w_{\mu}(\tau) = z_{\mu}(\tau)(1-\beta) + {\bar z}_{\mu}(\tau)\beta \;$ and also
$$ a_{\mu\nu} = \left ( \frac{\partial w_{\mu}}{\partial \beta}
\frac{\partial w_{\nu}}{\partial \tau} - \frac{\partial
w_{\mu}}{\partial\tau}\frac{\partial w_{\nu}}{\partial\beta} \right ) $$

Doing nonrelativistic expansion, it is easy to obtain:
 $$ \int d {\sigma}_{\mu\rho}(w)\>G_{\mu\rho}(w) =
\int\limits_{T_1}^{T_2}d\tau \int\limits_0^1 d\beta \left ( (\vec E \vec
r(\tau)) + (\vec H \vec L)\frac{2}{i\mu}\left (\beta - \frac{m_1}{m} \right
) \right ) $$
where
$ L_k =
{\epsilon}_{ijk}\>r_i\> \frac1i\>\frac{\partial}{\partial r_j} $
and $ \vec E $ and $ \vec H $ are chromoelectric and chromomagnetic
fields respectively, they are in the so called Feynman-Schwinger gauge here
( fixed point is $x_0$),
 and in this gauge $\>\Phi(x,x_0) = 1$.

Stricly speaking the last term in brackets is outside from our accuracy because
it is of the order of $\frac{v}{c} $. Deriving $G_2^A$ we have omitted all such
effects (the nonladder perturbative exchanges, spins of quarks et al.)
Nevertheless, there may be some special cases, for example
quarkonium-nucleus scattering via chromomagnetic fields, arizing from the
nucleons Fermi motion, where nontrivial contribution from the $\vec H$ - term
exists.

G.f. in momentum space is as follows:
\begin{eqnarray*}
G^A_2(p_{in}, p_{out}) = \int e^{-ip_{in}z_1}d^3\vec R_1 dT_1
e^{ip_{out}z_2}d^3\vec R_2 dT_2\cdot\nonumber \\
{\cdot}<out|G^A_2(z_1, z_2)|in>,
\end{eqnarray*}
where $z_i=(\vec R_i(T_i), T_i) ,\; i=1,2 $, and $|in>$ ; $|out>$
-are the asymptotic states of the singlet Hamiltonian.

Also after returning to the Minkowskii space one can define the "amputated"
G.f. (factor $ 2\mu $ is a consequence of our nonrelativistic
normalization ):
$$ {\hat G}_2^A(p_{in}, p_{out})
= 2\mu\cdot\>\lim_{p^2\to m^2}\>
(p_{in}^2-m_{in}^2)\>G_2^A(p_{in},p_{out})\>(p_{out}^2-m_{out}^2) $$

Using spectral decomposition
we obtain the expression
for the nonrelativistic G.f.
$ {{\hat G}}_2^A $
in terms of ${\vec E}^a$ operators:


\begin{eqnarray} &{\hat G}_2^A(p_{in},p_{in}+q) = 2m\cdot\left
(\frac{-g^2}{2N_c}\right )\int\frac{d^4k}{(2\pi)^4}\int\frac{d^3\vec
p}{(2\pi)^3}\>\int\limits_0^1d\beta \int\limits_0^1d\beta' \nonumber\\
&<out|{\vec r\>}'{\vec E}^a\left ( \frac{q}{2}-k\right )\;
 e^{-i{\vec r\>}'\left (\frac{m_2}{m} - {\beta}'\right )\left (
\frac{\vec q}{2}- \vec k\right )}\>|p>
\>\frac{2mi}{(p_{in}+\frac{q}{2}+k)^2 - m_p^2 + i0} \>\cdot \nonumber\\
&\cdot<p|{\vec r\>}{\vec E}^a\left (\frac{q}{2} + k\right )\>e^{-i{\vec
r}\left ( \frac{m_2}{m}-\beta\right)\left(\frac{\vec q}{2}+\vec
k\right)}|in> \end{eqnarray}
Taking into account $ {p_{in}=(m+\frac{m{\vec
v}^2}{2}+E_1^{in}\>,m\vec v\>)},\> $ for nonrelativistic regime and also
$ {m_1=m_2=\frac{m}{2}} $ one obtaines for $ q=0 $ :
 \begin{eqnarray} &{\hat G}_2^A =
2m\cdot\left (\frac{-g^2}{2N_c}\right
)\int\frac{d^4k}{(2\pi)^4}\int\frac{d^3\vec
p}{(2\pi)^3}\>\int\limits_0^1d\beta \int\limits_0^1d\beta' \nonumber\\
&<out|{\vec r\>}'{\vec E}^a(-k)\;
 e^{i\vec k{\vec r\>}'\left (\frac12 - {\beta}'\right )
  }\>|p>
\>\frac{i}{k_0 + E_1^{in} - E_8^p +i0} \>\cdot \nonumber\\
&\cdot<p|{\vec r\>}{\vec E}^a (k)\>e^{-i\vec k{\vec
r}\left ( \frac12-\beta\right) }|in> \end{eqnarray}

Up to a normalization factor $ {\hat
G}_2^A $ is a scattering amplitude here.
 Formulae (2), (3) generalize a well-known result from
\cite{peskin}, \cite{bhanot}. They have been obtained here in a
gauge-invariant way, starting from the quark G.f.
In an effort to demonstrate a nonabelian
essence of (2), (3) it should be allowed the necessity to average the
operator $ {\vec E}^a{\vec E}^a $ over the field source states.  Let us
define $ {{\vec E}^a(k) = \int d^4x e^{ikx}\>{\vec E}^a(x)} $ and
\begin{equation} \int
d\left(\frac{x+y}{2}\right)\>e^{iq\left(\frac{x+y}{2}\right)}\>
<g^2\>E_i^a(x)E_j^a(y)>_A\;=\;{\delta}_{ij}\>{\Delta}_q(x-y)
\end{equation}
For zero momentum transfer:
$$
{\Delta}_{q=0}(x-y)\>\equiv\> {\Delta}(x-y)
$$
The bilocal correlator ${\Delta(x-y) \equiv \Delta(\vec x - \vec y , x_4 - y_4)}$
is defined
in an Euclidean space, hence one needs in Wick rotation $
{k_0\to -ik_4\;} ;\;{\;it\to\tau} $:
\begin{eqnarray} &{\hat G}_2^A(q=0) =
2m\cdot\frac{1}{2N_c}\int\frac{dk_4}{2\pi}
\int\limits_0^1d\beta\int\limits_0^1d{\beta}'\nonumber\\
&<out|{{r}_i\,}'\;\frac{\Delta\left(
\left(\frac12-{\beta}\right){\vec r\>} - \left(\frac12-{\beta}'\right)
{\vec r\>}',\;k_4\right)}{-ik_4 +
E_1^{in} - H_8 + i0 }\;r_i\>|in>
\end{eqnarray}
where ${ {\Delta}(\vec x - \vec y, k_4)
= \> \int d(x_4-y_4)\>e^{ik_4(x_4-y_4)}\>\Delta(x-y)} $.
Is it possible to consider the operator in brackets as a local one , as one
often does? If  $\Delta(\vec r, k_4)$
falls as a function of $k_4$ on a distances about
 $(T_g)^{-1}$, characteristic values of $k_4$ in the integral are controlled by
${\mu}_T\equiv (T_g)^{-1}$.  Perturbative calculations give us $ E_n
= -\frac{\mu(C_F{\tilde{\alpha}}_s)^2}{2n^2} $, so for charmonium
$| \epsilon | \sim600\>Mev $ ($\epsilon$ is a bound energy here,
$\mu$ - reduced mass, for
$m_1=m_2=m/2$ one has $\mu=m/4$), ${\tilde{\alpha}}_s $
is an effective strong coupling constant - with radiative corrections of the
first order (see in \cite{quarkonia} ) .
Such distances are more than quarkonium size and more than
 $T_g$, which are known from the lattice calculations and heavy quarkonium
energy shifts is about $(1 Gev)^{-1}$. We will neglect $\vec x - \vec y$
in $\Delta(\vec x - \vec y, k_4)$, i.e. use approximation
$ exp\>\left(i\vec k \vec r \left(1/2 - \beta \right)\right)\approx 1$.

 Nevertheless, even so 'space locality' exists, (with the rising accuracy
if quarkonium radius decreases) , there is a nontrivial dependence on
$(x_4 - y_4)$, in the correlator.
A neglection of it is equivalent to the limit ${\mu}_T\to 0$, which we
 shall discuss later on. So, it is obvious the necessity of special analysis of
  $ {\vec E}(x){\vec E}(y) $ locality hypothesis both perturbative
 and nonperturbative cases.


Let us make such analysis for charmonium-nucleon scattering with low energy.
They usually use different parametrization for correlator
$ {\Delta}(x-y) $ ( \cite{quarkonia}, \cite{dig} ),
taking into account lattice
data or some models.
We propose to use two forms for
correlator $ {\Delta}(x-y) $:
$$ {\Delta}_1^h(z) =
{\Delta}_1^h(0)\;e^{-\frac{|z|}{T_g}}\;\;;\;\;\;\; {\Delta}_2^h(z) =
 {\Delta}_2^h(0)\;e^{-\frac{z^2}{{T_g}^2}} $$

The letter "$h$"
in $\Delta_{1,2}^h(z) $  reminds us, that the average here should be over
all field configurations inside the hadron (not over the vacuum
fields!), so there is a question at this point, does the structure of these 
correlators coincide with the vacuum one?  The positive answer to this 
question can be explained on the following way.  Hadron parameters in the 
stochastic vacuum model are caused by the vacuum structure and the cluster 
expansion of the gauge-invariant Wilson loop depends only on the 
integral moments of the field correlators, so they contain all information 
about QCD vacuum, which may be investigated by colourless objects 
{\cite{simonov88}}.  The way these moments are composed is not universal and 
it depends on the "hadron" (form of the loop, kinematics) but not the 
structure of correlators (see also \cite{dig}).  The only thing, that should 
be changed in our case is the normalization of the bilocal correlator.  In 
the chiral limit  (we use normalization condition from \cite{s}) :  
\begin{eqnarray} m_N = \> 
<h|{\Delta}{\Theta}_{\mu\mu}\>|h>&&\Delta{\Theta}_{\mu\mu} =
{\Theta}_{\mu\mu} - <{\Theta}_{\mu\mu}>_A \nonumber\\
{\Theta}_{\mu\mu}(x) =
\frac{\beta({\alpha}_s)}{4{\alpha}_s}\>G^a_{\mu\nu}(x)G^a_{\mu\nu}(x)&&
{\beta}({\alpha}_s) = -\frac{b}{2{\pi}}\>{\alpha}_s^2
\end{eqnarray}
where $ b=9 $ for three flavours. From (4) and (6) it is easy to obtain the
normalization condition for the correlator:  $$ {\Delta}_{1,2}^h(0) =
\left(\frac{2}{3}\right)^3{\pi}^2\cdot m_N $$
It is seen from the definition, that vacuum average of the correlators
has been already subtracted here,
so, we are indeed considering the scattering amplitude.

In such a manner, amputated G.f. is proportional to an amplitude, as already
noted. Normalized relativistic amplitude
$T$ is connected with ${\hat G}_2^A$ on a simple way
$ {T = 2m_N\cdot {\hat G}_2^A}$.  Then we get:
$$ T_1 =
2m_N\cdot\frac{2m}{2N_c}\>\int \frac{dk_4}{2\pi}\><out|{r_i}'\>
\frac{\Delta_1^h (0, k_4)}{-ik_4 - \Omega + i0}\>r_i|in> $$ where operator
 $\Omega = H_8 + |E_1^{in}| $. Also $$
 \Delta_1^h(0,k_4) = \int\limits_{-\infty}^{\infty} d\tau
e^{ik_4\tau}\Delta(0)\;e^{-\frac{|\tau|}{T_g}} $$

Finally,

 \vspace{0.5cm} $$ T_1 =
m\>{m_N}^2{\left(\frac{2}{3}\right)}^4\>{\pi}^2\;
<out|r_i\frac{1}{H_8 - E_1^{in} + {T_g}^{-1} - i0}r_i|in>
 \; ;
$$
Analogously for Gaussian correlator
${\Delta}_2^h(z)$ :  \begin{eqnarray} T_2 =
m\>{m_N}^2{\left(\frac{2}{3}\right)}^4\>{\pi}^2\;<out|r_i T_g \frac{\sqrt{\pi}}{2}\left(
1-{\Phi}\left(\frac{T_g(H_8-E_1^{in})}{2}\right)\right)\cdot\nonumber\\
\cdot e^{\frac{{T_g}^2(H_8-E_1^{in})^2}{4}}\>r_i|in>
\end{eqnarray}
here $ {\Phi}(x)=\frac2{\sqrt{\pi}}\int_0^x e^{-t^2}dt.\\ $

For numerical analysis we choose
$m=M_{{\eta}_c}=2980\>Mev,\; m_N = 938\>Mev$.
  The limit  $T_g\to\infty $ (so called Leutwyler-Voloshin limit)
leads to the regime the results of the paper {\cite{kaid}}  was obtained in which.
Indeed, it is easy to see, that using right value for the matrix element
 $ {<1s|r_i\frac{1}{H_8-E_1}r_i|1s>}
$,
$$ <1s|r_i\frac{1}{H_8-E_1}r_i|1s> =
\frac{7}{4}\>m{a^4}\;\;  $$ our results coincide:
 \begin{equation} T =
 7\>M_{{\eta}_c}m_c\cdot\left(\frac{2\pi}{9}m_Na^2\right)^2 \end{equation}

If one estimates the corresponding scattering length with the value
 $a$ from \cite{kaid}
${a = 0.8\; {(Gev)}^{-1} }$, which reproduces the quarkonium
bound energy for Coulomb wave functions, it is easy to obtain:
  $$ l =
 \frac{T}{8\pi(M_{{\eta}_c} + m_N)} = 0.012\>fm $$ instead $l = 0.05\>fm$ in
 \cite{kaid}. Taking into account the finite value of $T_g$, namely
  $(T_g)^{-1} = 1\>Gev$ one has:  $$ l_{1} =
 \frac{T_{1}}{8\pi(M_{{\eta}_c} + m_N)} = 7\cdot10^{-3}\>fm $$ $$ l_{2} =
\frac{T_{2}}{8\pi(M_{{\eta}_c} + m_N)} = 8\cdot10^{-3}\>fm $$
 This result means, that charmonium-nucleus bound states are possible only
for very heavy nuclei. Quarkonium-nucleon bound energy is so small, that
potential wall deep enough may be only for $A\sim200$.

 The above amplitudes equal to zero in another interesting case
$ T_g\rightarrow 0 \;  ( m_{N}$ is fixed).
In actuality, as it is
evident from the foregoing, nonperturbative interaction is of the different
type in this case.

As it is known, the system dynamics in this regime is natural to describe by
the local potential \cite{quarkonia} . It means, it would serve no purpose
to present the scattering in terms of gluon exchanges. In reality only bound
colourless object can propagate over large distances in the impact
parameter plane - it is reasonable to believe in this scheme, that it is a
reggeized glueball. This possibility was supported sometimes ago by glueball Regge
trajectory estimation \cite{pom} in the framework of MVC,
it seems close to the phenomenological one for the Pomeron. The next step
should be the Regge scattering description in terms of QCD
vacuum values, taking into account also direct nonperturbative interaction
according to \cite{nacht} and perturbative exchanges in the nonperturbative vacuum,
where they are
important.  This work is in progress now.

{\it\bf Acknowledgments}.
The author would like to thank Prof. Yu.A.Simonov for useful discussions and stimulating
influence and P.E.Volkovitsky for some useful comments.

\end{document}